\documentclass[multphys,vecphys]{svmult}

\usepackage{makeidx}     % allows index generation
\usepackage{graphicx}    % standard LaTeX graphics tool
\usepackage{multicol}    % used for the two-column index
\makeindex             % used for the subject index
\newcommand{\eq}{equation}

\makeatletter
\def\lsim{\mathrel{\mathpalette\gl@align<}}
\def\gsim{\mathrel{\mathpalette\gl@align>}}
\def\gl@align#1#2{\lower.6ex\vbox{\baselineskip\z@skip\lineskip\z@
    \ialign{$\m@th#1\hfil##\hfil$\crcr#2\crcr\sim\crcr}}}
\makeatother

\newcommand{\apj}{ApJ }
\newcommand{\apjl}{ApJL }

\newcommand{\prd}{Phys.\ Rev.\ D }
\newcommand{\aap}{A \& A }
\begin{document}

\title*{Effects of Small-Scale Fluctuations of Neutrino Flux in Supernova
Explosions}
% Use \titlerunning{Short Title} for an abbreviated version of
% your contribution title if the original one is too long
\author{Hideki Madokoro, Tetsuya Shimizu, \and Yuko Motizuki}
% Use \authorrunning{Short Title} for an abbreviated version of
% your contribution title if the original one is too long
\institute{RIKEN, Hirosawa 2-1, Wako 351-0198, Japan
\texttt{madokoro@postman.riken.go.jp}}
%
% Use the package "url.sty" to avoid
% problems with special characters
% used in your e-mail or web address
%
\maketitle

\abstract
We examine effects of small-scale fluctuations with angle in the neutrino
radiation in core-collapse supernova explosions.  As the mode number of
fluctuations increases, the results approach those of spherical explosion.
We conclude that global anisotropy of the neutrino radiation is the
most effective mechanism of increasing the explosion energy when
the total neutrino luminosity is given.

\section{Introduction}

It has been turned out that most simulations of core-collapse supernova
explosions with spherical symmetry fail to produce a successful
explosion\cite{Lie01}.  In addition, observations suggest that the ejecta of
core-collapse supernova explosions are aspherical(e.g., \cite{Wa02}).  These
facts lead us to multidimensional simulations.  In 2-D and 3-D simulations%
\cite{MiWiMa93,HeBeHiFrCo94,BuHaFr95,JaMue96,Me98,FrHe00,ShEbSaYa01,FrWa02,
KiPlJaMu03},  it has been shown that multidimensional effects, such as
convection inside the proto-neutron star and convective overturn around the
neutrino-heated region, increase the explosion energy and can trigger a
successful explosion\cite{HeBeHiFrCo94,JaMue96,KeJaMue96}.

When a proto-neutron star rotates, the neutrino flux is expected to be
enhanced along the rotational (polar) axis.  Janka and M\"onchmeyer%
\cite{JaMoe89a,JaMoe89b} first discussed the possibility of
aspherical neutrino emission from a rapidly rotating inner core.
They argued that a neutrino flux along the polar axis might become
three times greater than that on the equatorial plane.

Shimizu et al.\cite{ShYaSa94,ShEbSaYa01} proposed that the anisotropic
neutrino radiation should play a crucial role in the explosion mechanism
itself.  They carefully investigated the effects of anisotropic
neutrino radiation on the explosion energy.  They found that only
a few percent enhancement in the neutrino emission along the pole is
sufficient to increase the explosion energy by a large factor, and leads to
a successful explosion.  They also found that this effect saturates around
a certain degree of anisotropy.  It should be noted here that the assumed
rotational velocity of the inner core is very different between
Janka et al.\cite{JaMoe89a, JaMoe89b} and Shimizu et al.%
\cite{ShYaSa94,ShEbSaYa01}.

In the work of Shimizu et al.\cite{ShEbSaYa01}, they considered only a form
of global anisotropy; the maximum peak in the neutrino flux distribution was
located at the pole and the minimum at the equatorial plane.
On the other hand, Burrows et al.\cite{BuHaFr95} have
suggested that the neutrino flux can fluctuate with angle and time due to
gravitational oscillation on the surface of the proto-neutron star.
In this work, we introduce such small-scale fluctuations in the neutrino flux
in our numerical code by modifying the angular distribution of the neutrino
flux.  We aim to study the effects of these small-scale fluctuations on the
shock position, the explosion energy, and the asymmetric explosion.
Details are found in ref.\cite{MaShMo03}.

\section{Numerical Code}
We perform 2-D simulations by solving hydrodynamic equations
in spherical coordinates.  A generalized Roe's method is employed to solve the
hydrodynamic equations with general equations of state (EOSs). The details of
our numerical technique, together with the EOS and the initial condition used,
are described in the previous article\cite{ShEbSaYa01}.  In our study,
we have improved the numerical code of Shimizu et al.\cite{ShEbSaYa01};
the cells in the $\theta$-direction were shifted by half of the cell
size\cite{Sh95} in order to avoid a numerical error near the pole, although
the error was not serious for the investigation of the explosion energy.

In the present work, the local neutrino flux is assumed to be\cite{MaShMo03}
\begin{\eq}
  l_{\nu}(r,\theta) = \frac{7}{16}\sigma T_{\nu}^{4} c_{1}
  \left(1+c_{2}\cos^{2}(n_{\theta}\theta)\right)\frac{1}{r^{2}},
  \label{eqn:nuflux}
\end{\eq}

\noindent
where $\sigma$ is the Boltzmann constant, and $T_{\nu}$ is the temperature on
the neutrinosphere.  In equation (\ref{eqn:nuflux}), $n_{\theta}$ represents
the number of waves in the $\theta$-direction.  The case of $n_{\theta}=1$
corresponds to the global anisotropy, namely, no fluctuation.  $c_{2}$ is a
parameter which is related to the degree of anisotropy in the neutrino
radiation.  We see in
equation (\ref{eqn:nuflux}) that the neutrino fluxes in the $x$ (equatorial)
and $z$ (polar) directions become $l_{x} \equiv l_{\nu}(r,\theta=90^{\circ})
\propto c_{1}$ and $l_{z} \equiv l_{\nu}(r,\theta=0^{\circ})
\propto c_{1}(1+c_{2}),$ respectively.  The degree of anisotropy $l_{z}/l_{x}$
is then represented as
\begin{\eq}
  \frac{l_{z}}{l_{x}} = 1 + c_{2}.
  \label{eqn:anisotropy}
\end{\eq}

\noindent
The value of $c_{1}$ is calculated from $c_{2}$ and $n_{\theta}$ so as
to adjust the total neutrino flux to that in the spherical model at the
same $T_{\nu}$.

It should be noted here that the amplitude of fluctuations in the neutrino
flux distribution for an observer far from the neutrinosphere and that on
the neutrino-emitting surface are different.  When we observe the neutrino flux
far from the neutrinosphere, the local neutrino flux is seen as equation
(\ref{eqn:nuflux}).  On the other hand, the neutrino flux on the
neutrino-emitting surface has a profile similar to equation (\ref{eqn:nuflux})
but the amplitude is different.  In the latter, $c_{2}$ is replaced by $a$
where $a$ is a parameter which represents the degree of anisotropy of
the neutrino flux on the neutrinosphere.  It is preferable
that we compare the results for the same value of $a$, since $a$ is more
directly related to explosion dynamics.  The value of $c_{2}$,
therefore, is calculated from a given $a$, depending on $n_{\theta}$.
Although it is difficult to calculate the exact relationship between $c_{2}$
and $a$, we can estimate it by assuming that the strength of the
neutrino flux on the neutrinosphere is approximated by a profile of step
function.  This makes it possible to relate the value of $c_{2}$
to the value of $a$ for each $n_{\theta}$.  For detail, see Madokoro et al.%
\cite{MaShMo03}.

We set $a=0.31$ in this work.  This value of $a$ is chosen in such a way that
the value of $l_{z}/l_{x}$ for the global model ($n_{\theta}=1$) becomes 1.10.
The values of $c_{2}$ for each fluctuation model ($n_{\theta}=3,5$) are
accordingly calculated.  These are summarized in Table~{\ref{tab:tab1}}.
The neutrino temperature on the neutrino-emitting surface $T_{\nu}$ is assumed
to be 4.70 MeV.  In our simulation, we have 500 nonuniform radial zones which
cover from 50 to 10000 km in radius.  For $\theta$, we have 62 uniform angular
zones from $\theta=0^{\circ}$ to $\theta=90^{\circ}$ with equatorial
symmetry for $n_{\theta}=1$ and $3$.  Note that we use a double value
of the angular zones for $n_{\theta}=5$, which is different from that used
in ref.\cite{MaShMo03}.

\section{Results}
\begin{figure}
\centering
\includegraphics[height=5cm]{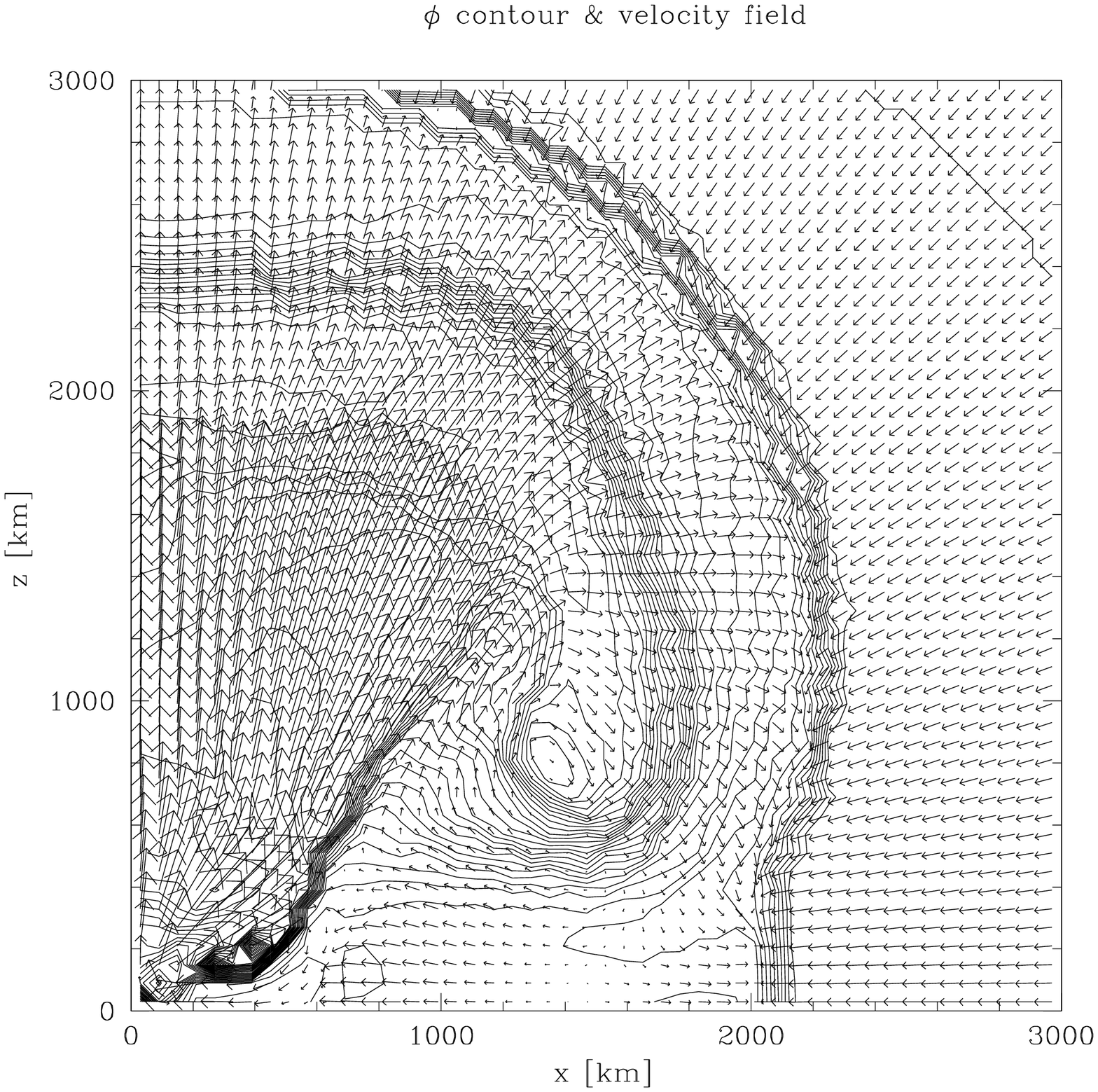}
\includegraphics[height=5cm]{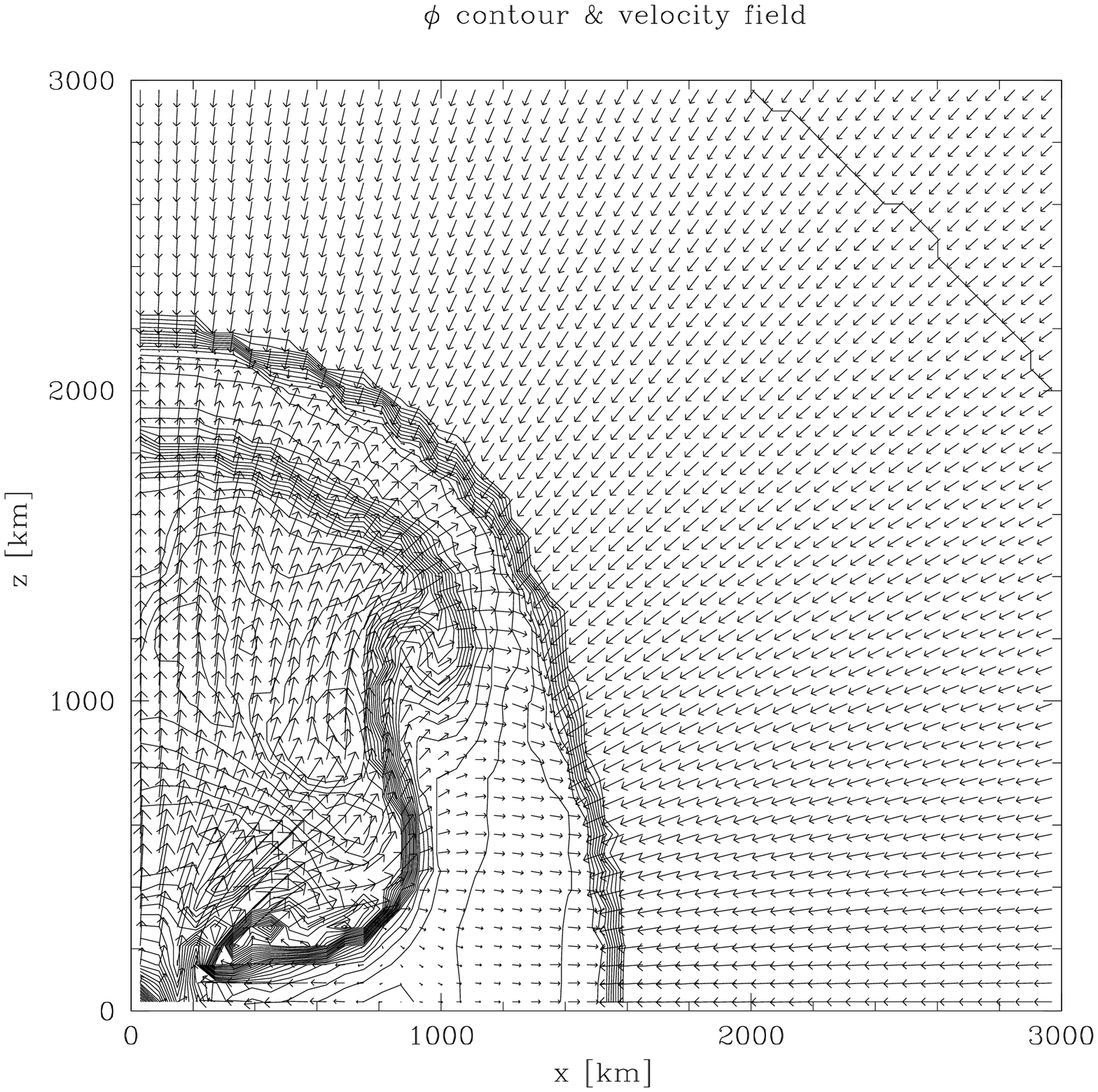}
\includegraphics[height=5cm]{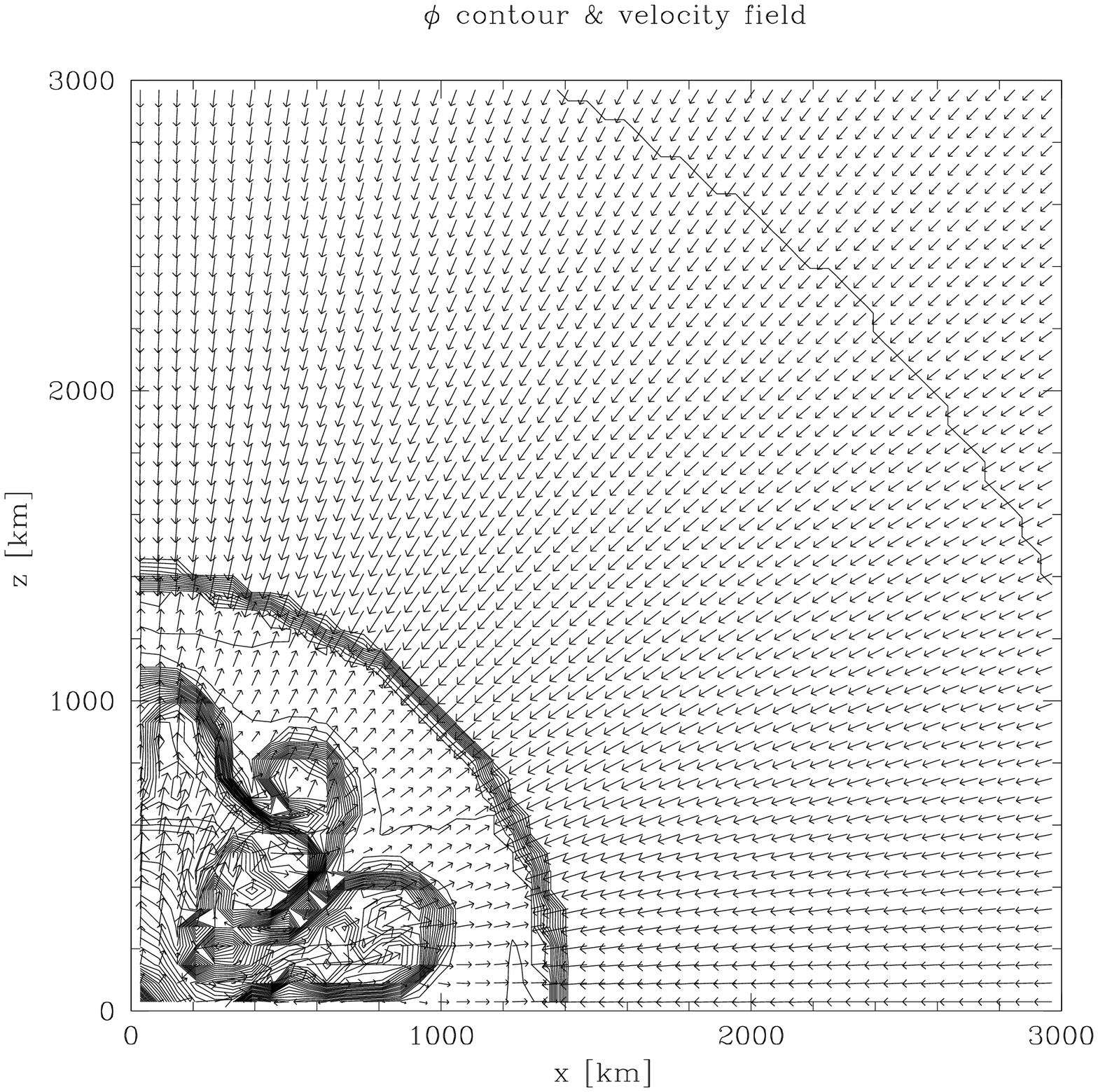}
\caption{Contour maps of the dimensionless entropy distribution and
           the velocity fields for the three models;
           top left: $n_{\theta}=1$ (model A1-T470) at $t=244$ms,
           top right: $n_{\theta}=3$ (model A3-T470) at $t=254$ms,
           bottom: $n_{\theta}=5$ (model A5-T470) at $t=252$ms.}
\label{fig:fig1}       % Give a unique label
\end{figure}

Figure~\ref{fig:fig1} depicts the contour maps of the dimensionless entropy%
\cite{ShEbSaYa01} distribution with the velocity fields for the three models
at $t\sim 250$ ms after the shock stall.  The shock front is represented by
the crowded contour lines at $r\sim 2000-3000$ km for the model A1-T470,
$r\sim 1600-2200$km for the model A3-T470, and $r\sim 1400$km for the model
A5-T470.  We see in Figure \ref{fig:fig1} that the shock front is largely
distorted in a prolate form for the model A1-T470.  This is because the
neutrino heating along the pole is more intensive than that on the equatorial
plane.  Due to increased pressure in the locally heated matter near the polar
axis, the shock front along the pole is pushed up, resulting in a prolate
deformation.  We find that the shock position becomes less extended than that
of the global anisotropy, and the shock front approaches the spherical shape
when the mode number of fluctuation increases.  This trend is especially
remarkable for the model A5-T470 in which the shock front is almost spherical.

\begin{figure}
\centering
\includegraphics[height=5cm]{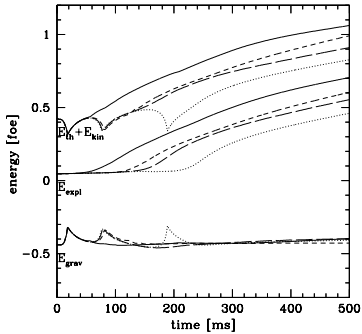}
\caption{Evolution of thermal and kinetic energy
           ($E_{\rm th}+E_{\rm kin}$), gravitational
           energy ($E_{\rm grav}$) and explosion energy ($E_{\rm expl}$)
           for the three models;
           solid line: $n_{\theta}=1$, short-dashed
           line: $n_{\theta}=3$, long-dashed line: $n_{\theta}=5$.
           For comparison, the result of the spherical explosion at the same
           $T_{\nu}$ is also plotted (dotted line).}
\label{fig:fig2}       % Give a unique label
\end{figure}

Figure~\ref{fig:fig2} shows the evolution of the explosion energy, as
well as the thermal, kinetic, and gravitational energies for the three models.
The energy gain for the case of $n_{\theta}=1$ is the highest among others at
all stages of the explosion.  It is also seen in Figure \ref{fig:fig2} that
the explosion energy decreases as the mode number of fluctuations in the
neutrino flux increases and finally approaches that of the spherical explosion.

Thus, we found that there are remarkable differences in the explosion energy
depending on the mode number of the fluctuations.  We also found that larger
number of modes in the fluctuations makes the result closer to that of
the spherical explosion.
This is because any small-scale fluctuations on the neutrinosphere are greatly
averaged out when the neutrino emission is observed far enough from the
neutrino-emitting surface.  Moreover, we found that a certain broad space
is needed to be heated by neutrinos to revive the stalled shock wave rigorously
and hence the global anisotropy ($n_{\theta}=1$) is the most effective
to increase the explosion energy.  These results support the claim
by Shimizu et al.\cite{ShEbSaYa01}.

%Burrows et al.\cite{BuHaFr95} suggested that the neutrino flux can fluctuate
%not only with angle but with time.  Such time fluctuations are, however,
%expected to reduce further the efficiency of anisotropy.

\section{Conclusion}
We have investigated the effects of small-scale fluctuations in the neutrino
flux on the core-collapse supernova explosion.  The profile of the neutrino
radiation field was specified taking its geometric effects into account.
Since the small-scale fluctuations are averaged out for radiative and
hydrodynamic reasons, the results of the fluctuation models become closer to
that of the spherical explosion.  Consequently, the global anisotropy is the
most effective mechanism in increasing the explosion energy when $L_{\nu}$
is given.  This supports the claim made by Shimizu et al.\cite{ShEbSaYa01}.

%
% For tables use
%
\begin{table}
\centering
%
% For LaTeX tables use
%
\begin{tabular}{|c|c|c|}
\hline
  Model & $n_{\theta}$ & $c_{2}$ \\
\hline
\hline
  A1-T470 & 1 & 0.100 \\
\hline
  A3-T470 & 3 & 0.051 \\
\hline
  A5-T470 & 5 & 0.035 \\
\hline
\end{tabular}
\caption{Simulated models.}
\label{tab:tab1}
\end{table}

%
%
% For figures use
%
%
%
% For figures use
%

% Non-BibTeX users please follow the syntax
% the syntax of "referenc.tex" for your own citations
%%%%%%%%%%%%%%%%%%%%%%%% referenc.tex %%%%%%%%%%%%%%%%%%%%%%%%%%%%%%
% sample references
% "physics"
%
% Use this file as a template for your own input.
%
%%%%%%%%%%%%%%%%%%%%%%%% Springer-Verlag %%%%%%%%%%%%%%%%%%%%%%%%%%

%
% BibTeX users please use
% \bibliographystyle{}
% \bibliography{}

\begin{thebibliography}{99.}
%
% and use \bibitem to create references.
%
% Use the following syntax and markup for your references
%
\bibitem{Lie01} M.~Liebendoerfer, A.~Mezzacappa,
  F.~-K.~Thielemann, O.~E.~B.~Messer, W.~R.~Hix, \& S.~W.~Bruenn:
  \prd \textbf{63}, 103004 (2001)
\bibitem{Wa02} L.~Wang et al: \apj \textbf{579}, 671 (2002)
\bibitem{MiWiMa93} D.~S.~Miller, J.~R.~Wilson, \& R.~W.~Mayle:
  \apj \textbf{415}, 278 (1993)
\bibitem{HeBeHiFrCo94} M.~Herant, W.~Benz, W.~R.~Hix, C.~L.~Fryer,
  \& S.~A.~Colgate: \apj \textbf{435}, 339 (1994)
\bibitem{BuHaFr95} A.~Burrows, J.~Hayes, \& B.~A.~Fryxell:
  \apj \textbf{450}, 830 (1995)
\bibitem{JaMue96} H.~-T.~Janka, \& E.~M\"uller: \aap \textbf{306}, 167 (1996)
\bibitem{Me98} A.~Mezzacappa, A.~C.~Calder, S.~W.~Bruenn, J.~M.~Blondin,
  M.~W.~Guidry, M.~R.~Strayer, \& A.~S.~Umar: \apj \textbf{495}, 911 (1998)
\bibitem{FrHe00} C.~L.~Fryer, \& A.~Heger: \apj \textbf{541}, 1033 (2000)
\bibitem{ShEbSaYa01} T.~M.~Shimizu, T.~Ebisuzaki, K.~Sato, \& S.~Yamada:
  \apj \textbf{552}, 756 (2001)
\bibitem{FrWa02} C.~L.~Fryer, \& M.~S.~Warren: \apjl \textbf{574}, L65 (2002)
\bibitem{KiPlJaMu03} K.~Kifonidis, T.~Plewa, H.~-T.~Janka, \&  E.~M\"uller:
  \aap, submitted (astro-ph/0302239) (2003)
\bibitem{KeJaMue96} W.~Keil, H.~-T.~Janka, \& E.~M\"uller:
  \apj \textbf{473}, L111 (1996)
\bibitem{JaMoe89a} H.~-T.~Janka, \& R.~M\"onchmeyer:
  \aap \textbf{209}, L5 (1989)
\bibitem{JaMoe89b} H.~-T.~Janka, \& R.~M\"onchmeyer:
  \aap \textbf{226}, 69 (1989)
\bibitem{ShYaSa94} T.~Shimizu, S.~Yamada, \& K.~Sato:
  \apj \textbf{432}, L119 (1994)
\bibitem{MaShMo03} H.~Madokoro, T.~Shimizu, \& Y.~Motizuki:
  \apj \textbf{592}, 1035 (2003)
\bibitem{Sh95} T.~M.~Shimizu: Ph.D.thesis, Univ. Tokyo (1995)
\end{thebibliography}
%
% Non-BibTeX users please use

%%%%%%%%%%%%%%%%%%%%%%%%%%%%%%%%%%%%%%%%%%%%%%%%%%%%%%%%%%%%%%%%%%%%%%

%%%%%%%%%%%%%%%%%%%%%%%%%%%%%%%%%%%%%%%%%%%%%%%%%%%%%%%%%%%%%%%%%%%%%%

\printindex
\end{document}